\documentclass[runningheads]{llncs}
\usepackage{graphicx}
\usepackage{amsmath}
\usepackage[colorlinks=true, urlcolor=blue, citecolor=blue]{hyperref}
\usepackage{booktabs}
\usepackage{cite}
\usepackage{tabularx}
\usepackage{xcolor}
\usepackage{lineno} % Include the lineno package
\definecolor{mygreen}{rgb}{0, 0.5, 0}

\title{Between Tool and Trouble: Student Attitudes Toward AI in Programming Education}
 
\titlerunning{Between tool and trouble}

\author{Sergio Rojas-Galeano $^1$, Julian Tejada $^2$, Fernando Marmolejo-Ramos $^3$}%\inst{1}}
\authorrunning{Rojas-Galeano et al}

\institute{$^1$ Universidad Distrital Francisco José de Caldas, Colombia\\
\email{srojas@udistrital.edu.co}\\[.2cm]
$^2$ Universidade Federal de Sergipe, Brazil\\
\email{julian@academico.ufs.br}\\[.2cm]
$^3$ Flinders University, Adelaide, Australia\\
\email{fernando.marmolejoramos@flinders.edu.au}}

\begin{document}

\maketitle

\begin{abstract}
This study examines how AI code assistants shape novice programmers’ experiences during a two-part exam in an introductory programming course. In the first part, students completed a programming task with access to AI support; in the second, they extended their solutions without AI. We collected Likert-scale and open-ended responses from 20 students to evaluate their perceptions and challenges. Findings suggest that AI tools were perceived as helpful for understanding code and increasing confidence, particularly during initial development. However, students reported difficulties transferring knowledge to unaided tasks, revealing possible overreliance and gaps in conceptual understanding. These insights highlight the need for pedagogical strategies that integrate AI meaningfully while reinforcing foundational programming skills.

\end{abstract}

\section{Introduction}

% \url{https://www.geekwire.com/2025/coding-is-dead-uw-computer-science-program-rethinks-curriculum-for-the-ai-era/?utm_source=Live+Audience&utm_campaign=78a0a94c89-nature-briefing-ai-robotics-20250715&utm_medium=email&utm_term=0_-b08e196e33-52592780}

The rapid advancement of AI conversational chatbots —such as ChatGPT\cite{openai2023gpt4} or Perplexity— has driven transformative changes across many fields\cite{bubeck2023sparks}, particularly in programming and coding generation. When deployed as AI-powered code assistants these models can generate, debug, and explain code with increasing fluency. Their growing presence in professional and educational settings raises important questions about their role in computer science education.

Integrating AI coding assistants into programming curricula presents both opportunities and challenges. On the one hand, they can enhance learning by offering instant feedback, generating example solutions, and assisting with common syntactic issues. On the other, there are concerns about overreliance, reduced engagement with fundamental concepts, and weakened development of core programming skills.

Previous research in educational technology has consistently shown that the effectiveness of technological tools depends on how they are pedagogically implemented. Studies involving intelligent tutoring systems and automated feedback mechanisms indicate that these tools can enhance learning when used to complement rather than replace traditional instruction.

However, the role of AI code assistants in programming education remains insufficiently understood. While most existing research has focused on professional development contexts or examined AI tools in isolation, there is a growing need to explore how students perceive and engage with these tools in educational settings. Understanding these perceptions is a crucial step toward assessing the broader pedagogical implications of AI integration.

This study contributes to addressing that gap by examining student attitudes and experiences with AI assistance during a programming assignment. While not intended to measure long-term learning outcomes, the research offers preliminary insights into how students perceive the benefits and limitations of AI tools, and how such perceptions may influence their learning practices, confidence and skill development in programming.

%\subsection*{Related Work}

Recent research has actively explored the impact of LLMs on programming education for first-year college students. Alves and Cipriano~\cite{alves2024give} analyzed student interactions with GPT models during programming assignments, including object-oriented programming tasks. Becker et al.~\cite{becker2023generative} investigated how tools like ChatGPT and GitHub Copilot compare to high-performing students in CS1, highlighting implications for automated support in introductory programming.

Fowler~\cite{fowler2024strategies} outlines some approaches to use LLMs for various assessment strategies in early programming education. Güner and Er~\cite{guner2025ai} conducted controlled experiments involving ChatGPT use, revealing different learner profiles among first-year students and how these profiles correlate with learning outcomes. Although they conducted a well-controlled within-subjects experiment comparing student performance with and without ChatGPT support, their study was situated in a structured lab setting with instructional interventions. They found that students benefited most when using AI for code refinement rather than full generation. Our study complements this work by focusing on students’ use and perceptions of AI in an authentic assessment context. By examining how novices interact with and reflect on AI assistance during real evaluative tasks, we address the gap between experimental findings and actual classroom practices.

Kulangara~\cite{kulangara2024designing} presented the design and evaluation of an LLM-supported educational platform aimed at teaching introductory programming, while Luo et al.~\cite{luo2025assessing} assessed personalized AI mentoring using GPT systems in computing education. Shynkar~\cite{shynkar2023influence} explored the effects of AI-generated code on novice programmers’ learning trajectories in higher education contexts.

Tian et al.~\cite{tian2024teaching} examined creative coding and conversational AI interactions across experience levels, providing insight into beginner engagement with tools like ChatGPT. Although Unlutabak et al.~\cite{unlutabak2023exploring} focused on material science education, their analysis includes perspectives on LLM applications relevant to novice programmers. Zhou et al.~\cite{zhou2024teachers} contributed an ethics-focused review, addressing LLM integration challenges in early computing courses from multiple stakeholder perspectives. This growing body of literature confirms that LLMs are reshaping the pedagogical landscape for first-year programming education, offering opportunities for both enhanced support and deeper reflection on foundational learning.

%\textcolor{orange}{\textit{Sergio, could you state here the goal of this study and its hypotheses?}}
%\textcolor{mygreen}{\textbf{Purpose and Hypothesis.}  
The purpose of this study is to investigate how students’ perceptions of AI-powered coding assistants influence their development of fundamental skills for the conceptual understanding of programming. We focus on the role these tools play in shaping students’ engagement across aided and unaided programming tasks. We hypothesize that students who perceive AI assistants primarily as \textit{answer providers} are more likely to exhibit superficial engagement and report weaker conceptual understanding, whereas those who view them as \textit{learning facilitators} or \textit{debugging aids} are more likely to demonstrate reflective use and deeper comprehension of core programming constructs. To explore this hypothesis, we analysed students’ self-reported experiences collected through an anonymous post-exam survey, focusing on their expressed purposes for using AI tools, perceived understanding of the code they submitted, and their ability to solve related tasks without AI assistance in a follow-up assessment.
%}

% \textcolor{lightgray}{This study has the purpose to investigate the usage, perceptions, and attitudes of students toward AI-powered code assistants within a controlled educational context. It is hypothesised that...\\
% Coloco una par de ideas e hipotesis aca PERO deja solo aquellas queue realmente see respondencon los datos:\\
% = Students will demonstrate differentiated usage patterns of AI coding assistants based on task complexity, with higher reliance on AI tools for syntactic debugging and lower reliance for conceptual problem-solving tasks (Hypothesis 1: Usage Patterns and Task Complexity).\\
% = Students who use AI coding assistants extensively during programming assignments will report lower self-efficacy in independent problem-solving compared to students who use AI tools sparingly or for specific purposes only (Hypothesis 2: Confidence and Self-Efficacy).\\
% = Students who perceive AI coding assistants primarily as ``answer providers" will demonstrate weaker conceptual understanding of programming fundamentals compared to students who perceive these tools as ``learning facilitators" or ``debugging aids." (Hypothesis 3: Learning Approach and Conceptual Understanding)\\
% This can be tested by:
% -Surveying students about their perceptions of AI tool purposes\\
% -Conducting follow-up assessments without AI assistance\\
% -Analyzing the quality of student explanations of their code\\
% -Comparing performance on conceptual questions vs. implementation tasks}

\section{Methodology}

\subsection{Participants}
The research was conducted with a group of 20 undergraduate students registered in an introductory Java-based Object-Oriented Programming (OOP) course, offered by the Computer Science department at the District University of Bogotá, in the first semester of 2025. All participants had prior exposure to programming concepts, but varied in their experience with AI code assistants. The study adhered to ethical principles outlined in the Helsinki Declaration \cite{wen2025helsinki}, and no personal or sensitive information was gathered.

\subsection{Procedure}

The experiment took place during three class periods across three successive weeks in the department's computer lab. Each participant attended the sessions in the following sequence:

\paragraph{Session 1: Programming with AI assistance.}

In the first session, students were assigned to develop a complete Java application from scratch based on a medium-difficulty specification (see supplementary materials). The task involved building an object-oriented system for managing a movie theater billboard, incorporating inheritance, dialog-based user interfaces, and collections. Students had access to the BlueJ IDE\cite{kolling2003bluej}, the official Java API documentation, and were explicitly allowed to use any AI-powered code assistant of their choice (e.g., ChatGPT, Perplexity, Grok, DeepSeek). The session lasted 90 minutes and was conducted individually under supervision.

\paragraph{Session 2: Programming without AI assistance.}

One week after the initial session, students returned for a second session focused on extending their original application. Each student received a randomly assigned programming task selected from a curated set of short, low-difficulty extensions (see supplementary materials). These tasks involved operations such as traversing object lists, applying conditional logic, and updating the state of specific objects. Implementation required a small adjustment to the existing interface—for example, by adding a new button to trigger the added functionality.

During this session, students were not allowed to use any AI assistance. They worked individually with access only to the BlueJ IDE and the official Java API documentation. To ensure compliance, all computers were offline and students were instructed to place their mobile phones face down on the desk. The session was supervised and lasted 30 minutes.

\paragraph{Session 3: Post-Experiment Survey.}
In the following class session, students completed an anonymous computer-based survey designed to capture their experiences and perceptions about AI use in programming tasks. The survey included Likert-scale items and open-ended questions. 

\subsection{Materials}
Participants underwent both main experimental conditions (coding with AI assistance and without it). During session 3, they responded to sets of Likert scale questions and open-ended questions aimed at assessing their views and attitudes about AI, including perceived usefulness, conceptual support, and self-reported confidence levels (see supplementary materials). The survey comprised:

%No direct measures of program correctness or completeness were collected; instead, the emphasis was on subjective student experiences and behavioural responses.

%\subsection{Data Collection}

\begin{itemize}
    \item[\textbullet] \textbf{Multiple choice and Likert-scale items} assessing students’ programming proficiency, AI tool usage, and perceived effects of AI on task-solving (e.g., coding speed-up, concept understanding, solution space exploration).
    \item[\textbullet] \textbf{Open-ended questions} exploring students’ experiences in greater depth, including their understanding and adaptation of AI-generated code, perceived difficulties, general views of the exam, and how confident they felt with and without AI assistance.
\end{itemize}

\subsection{Statistical Analysis}
The survey's rating data were examined using techniques designed for categorical data analysis \cite{bilder2017analysis}. Specifically, we employed the likelihood ratio $G^2$ test through the \texttt{loglm} function in the \textbf{MASS} R package, supplemented with mosaic plots generated using the \texttt{mosaic} function from the \textbf{vcd} R package. The \texttt{loglm} function fits log-linear models to multi-way contingency tables using a method known as iterative proportional fitting \cite{haberman1972algorithm}. Three-way and two-way contingency tabular data were analysed.

Three-way contingency models were studied where one factor was the chosen rating (five possible levels: nothing, slightly, moderately, strongly, completely), another factor was the participant's programming proficiency (two levels: basic and average), and the third factor was the question being assessed (e.g., the usability question included four sub-questions); here we denote these factors as $[A]$, $[B]$, and $[C]$ respectively. 

Because \texttt{loglm} is a model-oriented function, we examined main effects along with two- and three-way interactions in three-way contingency models. Model 1 included only main effects (complete independence, i.e. $[A][B][C]$) and served as the baseline model, representing the most parsimonious structure. Models 2, 3, and 4 incorporated distinct two-way interactions: $[AB][C]$, $[AC][B]$, and $[BC][A]$, respectively. A significant two-way interaction in a log-linear model such as $[AC][B]$ indicates that $A$ and $C$ are associated independently of $B$ (i.e. the relationship between $A$ and $C$ remains consistent across levels of $B$). Model 5 represents the saturated model, including all main effects, all possible two-way interactions, and the only possible three-way interaction (i.e. $[ABC]$). The models' deviance ($G^2$ likelihood ratio statistic), degrees of freedom (df), $p$-value, and Akaike Information Criterion (AIC) are reported.

In the case of two-way contingency models, the analysis focused solely on the association between two factors. These factors were participants' programming proficiency (with two levels: basic and average) and the categorical response options available. These two-way contingency models were applied to questions 21 and 22 of the survey (see supplementary materials). A significance level of $p < 0.05$ was used in all categorical data analyses.

% \textcolor{red}{Julian aca demos explicar como interpretar las tablas con resultados de esa modelacion... e.g. `deviance', df, etc. Por ejemplo hay que especificar en la tabla cada modelo; e.g. modelo 1 = solo main effects, modelo 2 = two-way interactions entre que factor y que factor y cual queda fijo, modelo 3 = interaccion de 3 vias.}

Mosaic plots offer a visual representation of a contingency table by partitioning a rectangle into tiles, where each tile's area is proportional to the frequency of the cell. The width and height of these tiles correspond to the marginal distributions of two or more categorical variables, making deviations from independence and association patterns clearly visible \cite{hartigan1981mosaics}. Person residuals indicate instances where observed counts surpass expectations (i.e., \( r > 0 \)) or fall short (i.e., \( r < 0 \)), with darker colors representing larger deviations. As a general guideline, \( |r| \approx 1 \) typically suggests random noise, \( |r| \geq 2 \) indicates moderate evidence of lack of fit, and \( |r| \geq 4 \) signifies very strong evidence. Since residuals are approximately \( N(0, 1) \) under the model, these thresholds are comparable to \textit{z}-scores. In simpler terms, by examining the darkest tiles, we can identify which combinations of categories are primarily responsible for the overall association. This helps determine where follow-up investigations or design modifications could have the most significant impact.

% Large Language Models (LLMs) have shown promise in assisting with thematic and content analyses, though they necessitate careful supervision. Research indicates that LLMs, such as ChatGPT, can detect basic themes and patterns in qualitative datasets, enhancing the efficiency of the analysis process when combined with traditional methods \cite{morgan2023exploring}. Despite their limitations in fully grasping nuanced semantic meanings, LLMs can help identify recurring topics and concerns in textual data, thereby aiding researchers and enabling more thorough thematic analyses \cite{han2025can}. Drawing on this evidence, we prompted Gemini 2.5 Pro to conduct a thematic analysis on the responses to the open-ended questions. \textcolor{orange}{For this task we used the prompt: ``xxxxxx''. In addition, we requested the LLM to deliver qualitative insights concerning sets of open-ended questions using the prompt ``ccccc''.}

Open-ended responses were examined through text mining and natural language processing (NLP) techniques via the R packages \textbf{syuzhet} for sentiment analysis and \textbf{topicmodels} for topic modelling using Latent Dirichlet Allocation (LDA is an unsupervised learning method that identifies topics in a collection of documents by assuming that documents are mixtures of various latent topics \cite{blei2003latent}). This qualitative data was also analysed via content analysis (see supplementary files).

All the R codes and datasets used in the analyses described above are available at \url{https://cutt.ly/OrFdufkH}).

%%%%%%%%%%%%%%%%%%%%%%%%%%%%%%%%%%%%%%%%%%%%%%%%%%%%%%%%%%%%%%%%%%%%%%
%%%%%%%%%%%%%%%%%%%%%%%%%%%%%%%%%%%%%%%%%%%%%%%%%%%%%%%%%%%%%%%%%%%%%%
%%%%%%%%%%%%%%%%%%%%%%%%%%%%%%%%%%%%%%%%%%%%%%%%%%%%%%%%%%%%%%%%%%%%%%

\section{Results}

The results are structured by analysis type, starting with the findings from categorical data analysis of Likert-type questions from both sessions, followed by the presentation of text mining and NLP models applied to the open-ended question responses.

\subsection{Categorical data analyses}

\begin{figure}[h]
\centering
\includegraphics[width=1\textwidth]{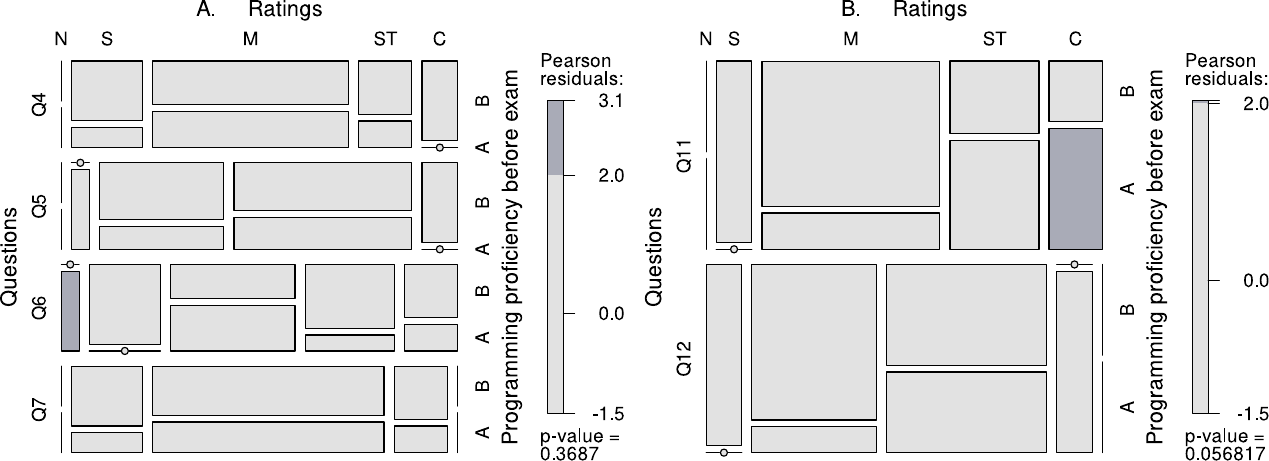}
\caption{Mosaic plot illustrating the relationship between `programming proficiency before the exam' (with levels basic (B) and average (A)), questions assessing the extent and effectiveness of AI usage (Q4: \textit{Level of AI usage during the exam}, Q5: \textit{Did AI help me solve the exercise faster in the first part of the exam?}, Q6: \textit{Did AI help me understand the concepts better in the first part of the exam?}, and Q7: \textit{Did AI help me explore different solutions in the first part of the exam?}) (plot A); questions assessing the perceived difficulty of the second session (Q11: \textit{In the second part of the exam, without the help of AI, were you able to solve the exercise?} and Q12: \textit{In terms of difficulty level and/or effort, do you consider that the second part of the exam was easier to solve than the first part?}) (plot B), and the associated ratings (nothing (N), slightly (S), moderately (M), strongly (ST), and completely (C)). }
\label{fig1}
\end{figure}

The results revealed non-significant associations in all models examined (with the exception of Models 1 and 4 in Table \ref{table1}). The significance of Model 1 suggests that at least one association exists among the variables, while Model 4 indicates a link between participants' programming proficiency and the assessed question, independent of rating choices. However, this finding is of limited interest, as rating choices—a key variable—do not interact in this model (see Figure \ref{fig1}).

\begin{figure}[h]
\centering
\includegraphics[width=1\textwidth]{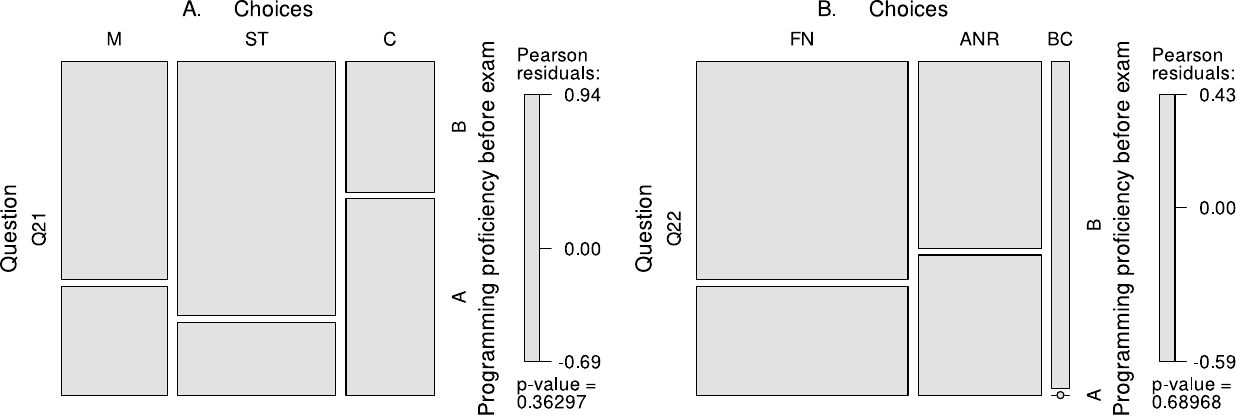}
\caption{Mosaic plot illustrating the relationship between `programming proficiency before the exam' (with levels basic (B) and average (A)) and response options to question 21 (\textit{Do you think AI is a useful tool for learning to program?}) (plot A) and question 22 (\textit{With the arrival of AI tools that automatically generate code, which of the following options best describes your opinion about learning to program?}). Rating options Q21: M: moderately, ST: strongly, and C: completely. Response choices Q22: FN: AI and Coding Fundamentals, ANR: Programming Aid, Not Replacement, BC: for Basic Code and Efficiency.  }
\label{fig2}
\end{figure}

The two-way contingency models indicated a non-significant association between programming proficiency and the rating options in question 21 ($G^2$(2)=1.99, $p$-value=0.36, AIC=27.90) and a non-significant association between programming proficiency and the multiple choices in question 22 ($G^2$(0)=0, $p$-value=1, AIC=27.46) (see Figure \ref{fig2}).

% \begin{figure}[h]
% \centering
% \includegraphics[width=1\textwidth]{img/figure2.pdf}
% \caption{Mosaic plot illustrating the relationship between `programming proficiency before the exam' (with levels basic (B) and average (A)), in A questions regarding perception of programming learning  (Q8: \textit{Understanding AI-generated code}, Q9: \textit{Code adaptation} and Q10: \textit{Confidence in AI-generated code }). In B, questions regarding the second part of the exam (without AI) (Q13: \textit{ Transfer to second part}, Q14: \textit{Confidence without AI }, Q15: \textit{Difficulties without AI }, Q16: \textit{Impact of AI on second part }, Q17: \textit{General perceptions of the exam format }, Q18: \textit{Comparison with and without AI }, Q19: \textit{Exam really assessed programming skills }, Q20: \textit{Suggestions to improve the exam })}. In C, questions related to expectations and preferences regarding the future use of AI in learning and programming processes (Q23: \textit{Future use of AI in programming}, Q24: \textit{AI role in programming courses}. UA:Understood and Adapted, UD:Understood but Doubts, PU:Partial Understanding/Difficult, LU:Limited Use/Errors, UsI:Used for Ideas/Assistance, ID: Increased Difficulty/Challenges, NS:No Significant Change/Preparedness, P-AI:Preference for AI/Missed AI, AI-T:AI as a Tool/Assistant, AI-L:AI for Learning/Teaching, CL: Cautious/Limited Integration, NO-AI:AI Should Be Avoided/Not Integrated )
% \label{fig2}
% \end{figure}

% \begin{table}[h]

% \caption{Hierarchical log-linear (Poisson) models with the factors rating choices $[A]$, participant's programming proficiency $[B]$ and the questions being assessed $[C]$. Model 1: $[A][B][C]$, model 2: $[BC][A]$, model 3: $[AB][C]$, model 4: $[AC][B]$, and model 5: $[ABC]$. }
% \label{tab:table1}
% \begin{tabular}{l l l l l}
% % \multicolumn{6}{c}{\textbf{Methods}}  
% \hline
% 	&	           $G^2$   	&	df 	&	$p$-value  	&	 AIC	\\
%     \hline
% \multicolumn{5}{c}{\textbf{Figure \ref{fig1}A}}			\\
% \hline
% Model 1   & 32.46 & 31 &        &          \\
% Model 2   & 32.46 & 28 & $-1.42e^{-14}$  &   \\
% Model 3   & 27.12 & 27 &  $5.34$  &         \\
% Model 4   & 14.25 & 19 &  $1.28e^{01}$  &       \\
% Model 5   &  0 &  0 &  $1.42e^{01}$  &        \\

% \hline
% \multicolumn{5}{c}{\textbf{Figure \ref{fig1}B}}	\\
% \hline
% Model 1   & 24.38 & 13 &        &           \\
% Model 2   & 24.38 & 12 &  0 &        1   \\
% Model 3   & 14.76 &  9 &  9.62 &        3    \\
% Model 4   & 13.13 &  9 &  1.63 &        0     \\
% Model 5   &  0 &  0 & 13.13 &        9    \\

% \hline

% \end{tabular}

% \end{table}

% \usepackage{graphicx}
\begin{table}[]
\centering
%\resizebox{\columnwidth}{!}{%
\begin{tabular}{lllll}
\hline
\multicolumn{1}{c}{\textbf{Model}} & \multicolumn{1}{c}{\textbf{$G^2$}} & \multicolumn{1}{c}{\textbf{df}} & \multicolumn{1}{c}{\textbf{$p$-value}} & \multicolumn{1}{c}{\textbf{AIC}} \\ \hline
\multicolumn{5}{c}{\textit{Figure 1A}}                                                                                                                                           \\ \hline
Model 1                            & 32.46                               & 31                               & 0.39                                    & 125.88                                \\
Model 2                            & 27.12                               & 27                               & 0.45                                    & 128.54                                \\
Model 3                            & 14.25                                & 19                               & 0.76                                    &  131.67                                \\
Model 4                            & 32.46                                & 28                               & 0.25                              & 131.88                                \\
Model 5                            & 0                               & 0                               & 1                                    &  155.41                               \\ \hline
\multicolumn{5}{c}{\textit{Figure 1B}}                                                                                                                                           \\ \hline
Model 1                            & 24.38                        & 13                              & 0.02                                      & 75.40                                \\
Model 2                            & 14.76                      & 9                               & 0.09                                         & 73.77                                \\
Model 3                            & 13.13                        & 9                             & 0.15                                          & 72.14                                \\
Model 4                            & 24.38                                 & 12                                      & 0.01                       & 77.40                                \\
Model 5                            & 0                               & 0                            & 1                                    & 77.01                  \\ \hline    \\         
\end{tabular}%
%}
\caption{Log-linear models with the factors `ratings' $[A]$, `participant's programming proficiency' $[B]$ and the questions being assessed $[C]$. Model 1: $[A][B][C]$, model 2: $[AB][C]$, model 3: $[AC][B]$, model 4: $[BC][A]$, and model 5: $[ABC]$. A higher $G^2$ suggests a significant difference between expected and observed data (low $p$-value). Low AICs signal better fits. dfs equal to 0 indicate no degrees of freedom left (all parameters estimated).}
\label{table1}
\end{table}

% \begin{table}[h]

% \caption{Hierarchical log-linear (Poisson) model resulting considering \textit{programming proficiency before exam} and \textit{AI usage level during exam} against AI perception variables from the second session: \textit{Solved without AI} and \textit{Part 2 easier than part 1}. The table showed and anova-like comparison of the three-way contingency models: independence model: no interactions (model 1), all two-way interactions but no three-way (model 2) and saturated (includes three-way)}
% \label{table2}
% \begin{tabular}{l l l l l l }
% % \multicolumn{6}{c}{\textbf{Methods}}  
% \hline
% 	&	           Deviance    	&	df 	&	Delta(Dev)  	&	 Delta(df)	&	P(\textgreater Delta(Dev))	\\
%     \hline
% \multicolumn{6}{c}{\textbf{Solved without AI}}			\\
% \hline
% Model 1   	&	20.416621	&	24	&	            	&	          	&		\\
% Model 2   	&	2.283265	&	9	&	18.133356	&	15	&	0.25572	\\
% Model 3   	&	0	&	0	&	2.283265	&	9	&	0.98616	\\
% Saturated 	&	0	&	0	&	0	&	0	&	1	\\
% \hline
% \multicolumn{6}{c}{\textbf{Part 2 easier than part 1}}	\\
% \hline
% Model 1   	&	22.361591	&	24	&	            	&	          	&	            	\\
% Model 2   	&	1.121668	&	9	&	21.239923	&	15	&	 0.12934	\\
% Model 3   	&	0	&	0	&	1.121668	&	9	&	0.99910	\\
% Saturated 	&	0	&	0	&	0	&	0	&	1	\\
% \hline

% \end{tabular}

% \end{table}

\subsection{Text mining, NLP, and content analyses}

The open-ended survey questions were organized into distinct thematic sets based on their focus. Questions 8-10 examined students' first session experiences, specifically how they interacted with, understood, and perceived the AI-generated code. Questions 13-20 explored the second session, comparing AI-assisted versus non-AI performance, overall exam experiences, and improvement suggestions. Questions 23 and 24 addressed future-oriented perspectives, capturing students' preferences and expectations for integrating AI into programming education and learning (see supplementary materials).

The LDA algorithm identified 2–4 latent themes within each question group, extracting the seven words with the highest $\beta$-weight (probability of appearing in that topic) for each theme. In the first group (Q8-Q10), two distinct topics emerged: troubleshooting code and conceptual clarity. This division reveals that respondents were split between addressing practical debugging issues and achieving higher-level conceptual understanding. The second group (Q13-Q20) yielded three topics: time pressure and iteration, coding help and comprehension, and data and results. These responses demonstrated broader thematic variety, encompassing workflow timing concerns, assistance requirements, and result interpretation challenges.
For Q23, two perspectives emerged: personal growth and practical application, encompassing themes such as learning gains and confidence building versus automation and project implementation. Finally, Q24 responses centered on two focal areas: future product evolution and ethical considerations, with topics including tool improvements and user expectations alongside academic integrity and originality concerns. 

The content analysis \cite{elo2008qualitative, nelson2020computational} identified several key patterns across the question sets. To this end, we used generative AI models to detect basic themes in the responses, complementing the expert analyst’s interpretation \cite{han2025can, morgan2023exploring}. Regarding students' first session AI experiences, most students successfully understood the AI-generated code, particularly those with prior programming knowledge. Active engagement was common, with students proactively seeking clarifications and adapting code to match their understanding. Code modification was widespread, as students refined variable names, simplified complex solutions, and extended AI output. However, confidence levels varied considerably—while many felt confident using AI-generated code, a substantial number harbored doubts about its efficiency or correctness. For the second session comparisons and overall experience, AI provided beneficial scaffolding for most students, offering foundational understanding and reusable code structures that facilitated their work. Confidence without AI was mixed; while many students felt prepared to work independently, syntax and language-specific issues emerged as major barriers. There was strong consensus that AI availability would have improved efficiency in the second part, though students recognized important learning trade-offs, acknowledging that while AI helps with speed, it may reduce skill development opportunities. Regarding exam evaluation, an overwhelming majority found the exam format effective for assessing programming skills. Students appreciated the novel approach but suggested timing improvements and better guidance. When considering future AI integration preferences, students favored a balanced approach, preferring AI in a support tool role for debugging, learning assistance, and doubt resolution. They emphasized selective use, wanting to maintain control while using AI for specific tasks. However, there was significant resistance to AI replacing core instruction, with many preferring minimal use or restricting AI to autonomous learning contexts rather than formal coursework, and some completely rejecting AI integration while emphasizing traditional pedagogy. The key insight emerging from thcis analysis is that students demonstrate a sophisticated understanding of AI as a tool that should supplement, not replace, fundamental learning processes (see details of this content analysis in the supplementary materials).

Finally, sentiment analysis revealed that responses to Q8-Q10 and Q13-Q20 exhibited neutral valence (median 0), with most comments containing a balanced mix of positive and negative words. Responses to Q23 (\textit{Future use of AI in programming}) demonstrated a more positive tone, while responses to Q24 (\textit{AI role in programming courses}) showed the highest level of positivity overall. The analysis also identified that a small subset of responses displayed enthusiastic sentiment across all question groups, regardless of the specific topic being addressed (see details of this NLP modelling in the supplementary materials).

%\textcolor{orange}{Lo que estoy pensando para esta sección es correr análisis usando NLP y text mining propiamente (por ejemplo podemos usar word clouds o cosas que se usen en el área) y decir al final en una frase ``(see supplementary files for an analysis of this data using content analysis)".}

%\textcolor{mygreen}{To complement the quantitative findings, students’ responses to the open-ended survey questions were analyzed using inductive content analysis \cite{elo2008qualitative, nelson2020computational}. This method involved systematically reviewing and coding the data to identify recurring themes and patterns, resulting in a set of emergent thematic categories for each question. The qualitative component offers deeper insight into students’ perceptions, experiences, and challenges with AI code assistants, with key themes emerging from their self-reported responses. For a detailed account of this analysis and representative quotes, please refer to the supplementary materials.}

%%%%%%%%%%%%%%%%%%%%%%%%%%%%%%%%%%%%%%%%%%%%%%%%%%%%%%%%%%%%%%%%%%%%%%%%%%%%%%%%%%%%
%%%%%%%%%%%%%%%%%%%%%%%%%%%%%%%%%%%%%%%%%%%%%%%%%%%%%%%%%%%%%%%%%%%%%%%%%%%%%%%%%%%%
%%%%%%%%%%%%%%%%%%%%%%%%%%%%%%%%%%%%%%%%%%%%%%%%%%%%%%%%%%%%%%%%%%%%%%%%%%%%%%%%%%%%

\section{Discussion}

This study explores the impact of novice programmers' perceptions of AI-powered coding assistants on their acquisition of foundational programming skills and conceptual knowledge. By analyzing student attitudes and experiences during a two-phase programming exam—first with AI assistance, then without—we assessed how these tools affect learning engagement and skill transfer. We hypothesized that students who primarily view AI assistants as answer providers would engage more superficially and demonstrate weaker conceptual understanding, whereas those who perceive them as learning aids or debugging tools would exhibit more reflective usage and deeper grasp of core programming concepts. Our findings partially confirm this hypothesis: although students generally found AI tools beneficial for code comprehension and confidence during initial development, they struggled significantly when applying knowledge without assistance, indicating patterns of overreliance and conceptual gaps. Due to the limited sample size, categorical data analyses yielded inconclusive results, but qualitative insights revealed that students who actively modified and adapted AI-generated code reported better understanding—though this did not consistently lead to stronger independent performance.

% This study offers preliminary insights into the role of generative AI assistants in programming education. While limited in scope, the combined quantitative and qualitative analysis helps illuminate how students interact with and learn from these tools.

In general, students found AI beneficial, especially for syntax and structural direction, in the initial session. Nevertheless, this assistance seemed mostly surface-level. Once AI tools were no longer available, a number of students encountered difficulties applying programming concepts on their own, indicating a limited transfer of skills. This trend suggests a potential \emph{cognitive offloading}, where reliance on AI may substitute rather than strengthen problem-solving skills. This concept is similar to `cognitive debt,' where continuous dependence on external systems, such as LLMs, could supplant the challenging cognitive processes necessary for independent thought \cite{kosmyna2025brain}.

The two-session design exposed critical gaps between AI-assisted performance and independent capability. While students showed initial confidence with AI support, they struggled with debugging and logic formulation when working alone, revealing weak internalized understanding despite strong assisted performance. These challenges were compounded by self-perception biases—students underreported both their AI reliance (rating it as \textit{Moderate} despite instructors observing widespread use) and their difficulties (calling the simpler second task \textit{Moderately} difficult while visibly struggling). Despite these challenges, students valued the exam format as educational, appreciating how it revealed both AI's benefits and limitations while emphasizing the need for balanced tool use that supports rather than replaces active reasoning and skill development.

% Confidence levels dropped in the unaided session. Despite initial optimism, students frequently expressed uncertainty with debugging and logic formulation when working alone. These challenges point to gaps in internalized understanding, even when initial AI-supported performance seemed strong.

% Discrepancies also emerged between self-perceptions and observed behavior. For example, while many students rated their AI usage as \emph{Moderate}, instructor observations indicated widespread and enthusiastic reliance. Similarly, students described the second task as \emph{Moderately} difficult, yet visibly struggled with it—despite its simpler structure. These contrasts suggest underreporting biases, perhaps due to social desirability or limited self-awareness.

% Nonetheless, students valued the exam format. Many described the two-session design as fair and educational, helping them recognize both the benefits and limitations of AI assistance. Several students highlighted the need to use AI tools in ways that support—not replace—their own reasoning and hands-on practice, favoring a balanced approach that encourages active engagement and skill development.

These findings underscore the need for thoughtful integration of AI tools in programming education and point toward broader implications in related domains such as statistics instruction, where both computational practices and conceptual reasoning are tightly interwoven. In introductory statistics courses that rely on R programming or leverage AI platforms like Julius AI, similar patterns may emerge. The risk of accumulating cognitive debt—where students rely on tool outputs without engaging in deeper analytical thinking—can be particularly salient. As \cite{wang2024scaffold} observed, while 85\% of STEM students report using generative AI tools for coursework, over half tend to input problems directly and rely on generated solutions, potentially bypassing critical learning processes.

% \textcolor{lightgray}{These findings underscore the need for thoughtful integration of AI into programming education, with particular implications for statistics instruction where both statistical reasoning and computational skills are essential. In statistics courses utilizing R programming or AI platforms like Julius AI, the risk of cognitive debt becomes especially pronounced. As ~\cite{wang2024scaffold} found, while 85\% of STEM students use generative AI tools for various tasks, over half simply input problems for AI to generate solutions, potentially bypassing critical learning processes. 
% }

This observation aligns with our findings of surface-level engagement that arises when AI assistance is readily available. To mitigate the risk of cognitive debt in statistics education, instructors should design scaffolding activities to position AI not merely as a solution provider, but as a catalyst for deeper learning. For instance, students might be instructed to: (1) critically evaluate AI-generated statistical analyses, interrogating assumptions and methodological choices; \mbox{(2) adapt} AI-produced R code to address edge cases or explore alternative scenarios, reinforcing both computational reasoning and statistical understanding; \mbox{(3) translate} AI outputs into plain-language explanations, thereby ensuring conceptual clarity beyond procedural accuracy; and (4) compare multiple AI-generated approaches to the same problem, developing analytical judgment and discernment.

% \textcolor{lightgray}{This aligns with our observations of surface-level engagement when AI assistance is readily available.
% To minimize cognitive debt in statistics education, instructors should design scaffolding activities that leverage AI as a learning partner rather than a solution provider. For instance, students could be asked to: (1) critique and validate AI-generated statistical analyses, requiring them to understand underlying assumptions and methodological choices; (2) modify AI-provided R code to handle edge cases or alternative scenarios, fostering deeper comprehension of both programming logic and statistical concepts; (3) explain AI outputs in plain language, ensuring conceptual understanding beyond computational results; and (4) compare multiple AI-generated approaches to the same statistical problem, developing critical evaluation skills.
% }

The \textit{confidence perception} and \textit{efficiency perception} pathways identified by~\cite{zhang2025paradox} suggest that although AI tools can bolster students' immediate sense of problem-solving ability, they may also foster an illusion of competence. In statistics education, this overconfidence poses particular risks when students encounter authentic data analysis tasks without AI assistance. To mitigate this, assessment strategies should incorporate both AI-supported and independent components—much like our two-session design—to help students calibrate their perceived and actual proficiency.
Moreover, as~\cite{wang2024scaffold} found, students who engaged with AI as a scaffold rather than a shortcut showed stronger development in problem-solving skills. Translating this to statistics instruction, one effective approach could involve structured prompting activities: students first articulate their reasoning, then consult AI tools, and finally reflect on any discrepancies between their initial thinking and the AI’s suggestions. This metacognitive loop encourages the internalization of statistical reasoning and reduces reliance on external validation.

% \textcolor{lightgray}{The \textit{confidence perception} and \textit{efficiency perception} pathways identified by \cite{zhang2025paradox} suggest that while AI tools enhance students' immediate problem-solving confidence, this may create an illusion of competence. In statistics education, this false confidence could be particularly dangerous when students face real-world data analysis challenges without AI support. Therefore, assessment strategies should include both AI-assisted and unassisted components, similar to our two-session design, allowing students to recognize their actual competency levels.
% Furthermore, as \cite{wang2024scaffold} observed, students who used AI tools as scaffolds rather than direct solution generators demonstrated better problem-solving development. In statistics courses, this could be implemented through structured prompting exercises where students must first articulate their statistical reasoning before consulting AI, then reflect on how AI suggestions align with or challenge their initial approach. This metacognitive practice helps students internalize statistical thinking patterns rather than developing dependency on external validation.
% }

Lastly, although this study was conducted with a small sample, its mixed-methods approach offers meaningful insights into students’ experiences with and without AI assistance. Future research could expand on these findings by incorporating behavioral data, longitudinal tracking, and controlled comparisons to deepen our understanding of AI’s long-term impact on the acquisition of computational thinking skills.

\section{Conclusions}

As generative AI tools become increasingly embedded in programming education, understanding their impact on learning is both urgent and complex. This exploratory study offers preliminary evidence that, while students find AI support valuable and report increased confidence during assisted tasks, these perceptions do not always align with independent performance or comprehension. Discrepancies between self-reports and instructor observations suggest that learners may underestimate their reliance on AI or overestimate their own mastery.

Such findings underscore the importance of designing AI-integrated activities that not only assist but also challenge students to internalize concepts and apply them independently. Educational strategies should aim to leverage AI for scaffolding, while encouraging active engagement and critical thinking. Mixed-method approaches—combining survey, behavioral, and performance data—will be essential to evaluate the true pedagogical impact of AI assistance.

Moreover, incorporating AI into the programming classroom is not simply a matter of access or efficiency, but of pedagogy. The goal should not be to replace effort with automation, but to use automation wisely—to support, not supplant, the learning process. Blending AI assistance with opportunities for independent problem-solving may help strike this balance, offering a promising direction for future research. In this new era of AI coding tools, it is essential to rethink programming education through strategies that foster engagement and gradually shift responsibility to the learner. 

Ultimately, the student perceptions captured in this study present a compelling case for reimagining paths that facilitate acquiring and developing computational thinking skills. We fully support this new direction, particularly as AI coding assistants become increasingly prevalent, meaning that ongoing research will be essential to ensure their effective and beneficial integration into programming education.

% \section*{References}

\end{document}